\documentclass[aps,prd,twocolumn,showpacs,showkeys,amsmath,amssymb]{revtex4}
\usepackage{graphicx}
\usepackage{dcolumn}
\usepackage{bm}
\begin{document}
%=====================================================================================
%=====================================================================================
\title{The $I^GJ^{PC}=0^+1^{-+}$ Tetraquark State}
%=====================================================================================
%=====================================================================================
%
\author{Hua-Xing Chen$^{1,2,3}$}
\email{hxchen@rcnp.osaka-u.ac.jp}
\author{Atsushi Hosaka$^{2}$}
\email{hosaka@rcnp.osaka-u.ac.jp}
\author{Shi-Lin Zhu$^{1}$}
\email{zhusl@phy.pku.edu.cn}
\affiliation{$^1$Department of Physics, Peking University, Beijing
100871, China
\\ $^2$Research Center for Nuclear Physics, Osaka University,
Ibaraki 567--0047, Japan \\ $^3$State Key Laboratory of Nuclear
Physics and Technology, Peking University, Beijing 100871, China}
\begin{abstract}
We study the tetraquark state with $I^GJ^{PC}=0^+1^{-+}$ in the QCD
sum rule. We exhaust all possible flavor structures by using a
diquark-antidiquark construction and find that the flavor structure
$(\mathbf{\bar 3}\otimes \mathbf{\bar 6})\oplus(\mathbf{6}\otimes
\mathbf{3})$ is preferred. There are altogether four independent
currents which have the quark contents $q s \bar q \bar s$. By using
both the Shifman-Vainshtein-Zakharov (SVZ) sum rule and the finite
energy sum rule, these currents lead to mass estimates around $1.8 -
2.1$ GeV, where the uncertainty is due to the mixing of two single
currents. Its possible decay modes are $S$-wave $b_1(1235) \eta$ and
$b_1(1235) \eta^\prime$, and $P$-wave $KK$, $\eta \eta$, $\eta
\eta^\prime$ and $\eta^\prime \eta^\prime$, etc. The decay width is
around 150 MeV through a rough estimation.
\end{abstract}
\maketitle
\pagenumbering{arabic}

Manifestly exotic hadron states which are not reached by three
quarks for baryons and a quark-antiquark pair for mesons provide one
of the most important subjects in hadron physics. The confirmation
of their existence (or nonexistence) and the study of their
structure are of great importance for the understanding of strong
interaction dynamics at low energy~\cite{exotic}.

Quantum numbers can tell whether a hadron is exotic or not. For
instance baryons with strangeness $S=+1$ and mesons with $J^{PC} =
1^{-+}$ are such states. For the baryon sector, the pentaquark
$\Theta^+$ has been studied intensively since 2003
~\cite{Nakano:2003qx}. But the existence is still controversial. For
the meson sector, the $\pi_1$ mesons of $I^G J^{PC} = 1^- 1^{-+}$
are listed as manifestly exotic states in the PDG for some
time~\cite{Yao:2006px,experiments,Lu:2004yn}, and a lot of
theoretical considerations have been made~\cite{theory,isoscalar}.
So far, many of them are for the isovector $I=1$ states. In
principle, an isoscalar state is also possible, though not observed
experimentally~\cite{isoscalar}. We have performed the QCD sum rule
analyses of the light scalar mesons ($\sigma$, $\kappa$, $f_0$ and
$a_0$), $Y(2175)$ and $\pi_1$ mesons~\cite{Chen,Chen:2008qw}. All
our results are consistent with the experimental observations.
Encouraged by this, we would like to extend the QCD sum rule
analysis using tetraquark currents for these $I^GJ^{PC}=0^+1^{-+}$
states.

%
%%%%%%%%%%%%%%%%%%%%%%%%%%%%%%%%%%%%%%%%%%%%%%%%%%%%%%%%%%%%%%%%%%%%%%%%%%%%%%
%---------figure 6*6
\begin{figure}[hbt]
\begin{center}
\scalebox{0.6}{\includegraphics{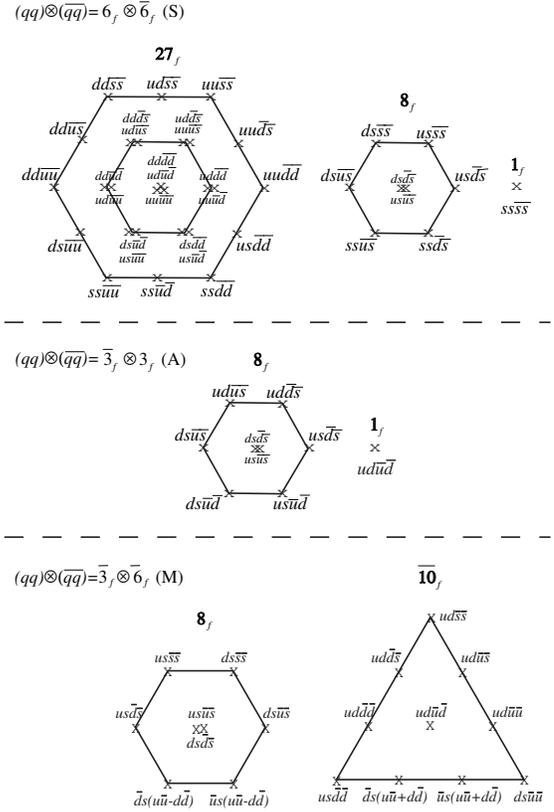}} \caption{Weight diagrams
for $\mathbf{6_f}\otimes\mathbf{\bar 6_f} (\mathbf{S})$ (top panel),
$\mathbf{\bar 3_f}\otimes\mathbf{3_f} (\mathbf{A})$ (middle panel),
and $\mathbf{\bar 3_f} \otimes \mathbf{\bar 6_f} (\mathbf{M})$
(bottom panel). The weight diagram for
$\mathbf{6_f}\otimes\mathbf{3_f} (\mathbf{M})$ is the
charge-conjugation transformation of the bottom one.}
\label{fig:tetra}
\end{center}
\end{figure}
%%%%%%%%%%%%%%%%%%%%%%%%%%%%%%%%%%%%%%%%%%%%%%%%%%%%%%%%%%%%%%%%%%%%%%%%%%%%%%
%

The QCD sum rule requires a computation of a two-point correlation
function in the form of operator product expansion (OPE), which is
then fitted by a phenomenological function to extract physical
hadron properties~\cite{sumrule}. To calculate the OPE, we need
employ an interpolating field (current) which couples to the
physical state we consider. For tetraquarks, there are several
independent currents and it is important to establish how one or
some of them should be chosen. We have systematically performed the
classification of currents by using the diquark-antidiquark
($(qq)(\bar q \bar q)$) construction~\cite{Chen,Chen:2008qw}. The
currents constructed from the quark-antiquark pairs ($(\bar q
q)(\bar q q)$) can be written as a combination of these ($(qq)(\bar
q \bar q)$) currents. We note here that the mixing can happen
between hybrid states, tetraquark states, and even six-quark states,
while the currents can also couple to all these states. However, it
always makes sense to clarify what a single channel problem tells us
before entering more sophisticated coupled channel problems.
Therefore, here we concentrate exclusively on the tetraquark
properties with some details.

The tetraquark currents with the quantum numbers $J^{PC} = 1^{-+}$
have been constructed in our previous paper~\cite{Chen:2008qw}.
Now we need construct the isoscalar ones. The flavor structures
are shown in Fig.~\ref{fig:tetra} in terms of $SU(3)$ weight
diagrams. The ideal mixing scheme is used since it is expected to
work well for hadrons except for the pseudoscalar mesons. In order
to have a definite charge-conjugation parity, the diquark and
antidiquark inside can have the same flavor symmetry, which is
either symmetric $\mathbf{6_f} \otimes \mathbf{\bar 6_f}$
($\mathbf{S}$) or antisymmetric $\mathbf{\bar 3_f} \otimes
\mathbf{3_f}$ ($\mathbf{A}$). Another option is the combination of
$\mathbf{\bar 3_f} \otimes \mathbf{\bar 6_f}$ and $\mathbf{6_f}
\otimes \mathbf{3_f}$ ($\mathbf{M}$), which can also have a
definite charge-conjugation parity.

From Fig.~\ref{fig:tetra}, we find that there are althgether six
isospin singlets:
\begin{eqnarray} \nonumber &&
q q \bar q \bar q (\mathbf{S})\, , q s\bar q \bar s (\mathbf{S}) \,
, s s \bar s \bar s (\mathbf{S}) \sim \mathbf{6_f} \otimes
\mathbf{\bar 6_f}~~~(\mathbf{S}) \, ,
\\ && q q \bar q \bar q (\mathbf{A}) \, , q s\bar q \bar s
(\mathbf{A}) \sim \mathbf{\bar 3_f} \otimes
\mathbf{3_f}~~~(\mathbf{A}) \, ,
\\ && \nonumber q s \bar q \bar s (\mathbf{M})
\sim (\mathbf{\bar 3_f} \otimes \mathbf{\bar 6_f}) \oplus
(\mathbf{6_f} \otimes \mathbf{3_f})~(\mathbf{M}) \, ,
\end{eqnarray}
where $q$ represents an $up$ or $down$ quark, and $s$ represents a
$strange$ quark. For each state, there are several independent
currents. We list them in the following.
\begin{enumerate}

\item For the three isospin
singlets of $\mathbf{6}_f \otimes \mathbf{\bar 6}_f$ ($\mathbf{S}$):
%
%%%%%%%%%%%%%%%%%%%%%%%%%%%%%%%%%%%%%%%%%%%%%%%%%%%%%%%%%%%%%%%%%%%%%%%%%%%%%%
\begin{eqnarray}\label{def:Scurrent}
&&
\begin{array}{ll}
\eta^S_{1\mu} & \sim u_a^T C \gamma_5 d_b (\bar{u}_a \gamma_\mu
\gamma_5 C \bar{d}_b^T + \bar{u}_b \gamma_\mu \gamma_5 C
\bar{d}_a^T) \\ & + u_a^T C \gamma_\mu \gamma_5 d_b (\bar{u}_a
\gamma_5 C \bar{d}_b^T + \bar{u}_b \gamma_5 C \bar{d}_a^T) \, ,
\\ \eta^S_{2\mu}
& \sim u_a^T C \gamma^\nu d_b (\bar{u}_a \sigma_{\mu\nu} C
\bar{d}_b^T - \bar{u}_b \sigma_{\mu\nu} C \bar{d}_a^T) \\ & + u_a^T
C \sigma_{\mu\nu} d_b (\bar{u}_a \gamma^\nu C \bar{d}_b^T -
\bar{u}_b \gamma^\nu C \bar{d}_a^T) \, ,
\end{array}
\\ &&
\begin{array}{ll} \eta^S_{3\mu}  & \sim u_a^T C \gamma_5 s_b
(\bar{u}_a \gamma_\mu \gamma_5 C \bar{s}_b^T + \bar{u}_b \gamma_\mu
\gamma_5 C \bar{s}_a^T) \\ & + u_a^T C \gamma_\mu \gamma_5 s_b
(\bar{u}_a \gamma_5 C \bar{s}_b^T + \bar{u}_b \gamma_5 C
\bar{s}_a^T) \, ,
\\ \eta^S_{4\mu}
& \sim u_a^T C \gamma^\nu s_b (\bar{u}_a \sigma_{\mu\nu} C
\bar{s}_b^T - \bar{u}_b \sigma_{\mu\nu} C \bar{s}_a^T) \\ & + u_a^T
C \sigma_{\mu\nu} s_b (\bar{u}_a \gamma^\nu C \bar{s}_b^T -
\bar{u}_b \gamma^\nu C \bar{s}_a^T) \, .
\end{array}
\\ &&
\begin{array}{ll} \eta^S_{5\mu} & \sim s_a^T C \gamma_5 s_b
(\bar{s}_a \gamma_\mu \gamma_5 C \bar{s}_b^T + \bar{s}_b \gamma_\mu
\gamma_5 C \bar{s}_a^T) \\ & + s_a^T C \gamma_\mu \gamma_5 s_b
(\bar{s}_a \gamma_5 C \bar{s}_b^T + \bar{s}_b \gamma_5 C
\bar{s}_a^T) \, ,
\\ \eta^S_{6\mu}
& \sim s_a^T C \gamma^\nu s_b (\bar{s}_a \sigma_{\mu\nu} C
\bar{s}_b^T - \bar{s}_b \sigma_{\mu\nu} C \bar{s}_a^T) \\ & + s_a^T
C \sigma_{\mu\nu} s_b (\bar{s}_a \gamma^\nu C \bar{s}_b^T -
\bar{s}_b \gamma^\nu C \bar{s}_a^T) \, .
\end{array}
\end{eqnarray}
%%%%%%%%%%%%%%%%%%%%%%%%%%%%%%%%%%%%%%%%%%%%%%%%%%%%%%%%%%%%%%%%%%%%%%%%%%%%%%
%
where $\eta^S_{1\mu}$ and $\eta^S_{2\mu}$ are the two independent
currents containing only light flavors; $\eta^S_{3\mu}$ and
$\eta^S_{4\mu}$ are the two independent ones containing one $s \bar
s$ pair; $\eta^S_{5\mu}$ and $\eta^S_{6\mu}$ are the two independent
ones containing two $s \bar s$ pairs.

\item For the two isospin
singlets of $\mathbf{\bar 3}_f \otimes \mathbf{3}_f$ ($\mathbf{A}$):
%
%%%%%%%%%%%%%%%%%%%%%%%%%%%%%%%%%%%%%%%%%%%%%%%%%%%%%%%%%%%%%%%%%%%%%%%%%%%%%%
\begin{eqnarray}\label{def:Acurrent}
&&
\begin{array}{ll}
\eta^A_{1\mu} & \sim u_a^T C \gamma_5 d_b (\bar{u}_a \gamma_\mu
\gamma_5 C \bar{d}_b^T -\bar{u}_b \gamma_\mu \gamma_5 C \bar{d}_a^T)
\\ & + u_a^T C \gamma_\mu \gamma_5 d_b (\bar{u}_a \gamma_5 C
\bar{d}_b^T - \bar{u}_b \gamma_5 C \bar{d}_a^T) \, ,
\\ \eta^A_{2\mu}
& \sim u_a^T C \gamma^\nu d_b (\bar{u}_a \sigma_{\mu\nu} C
\bar{d}_b^T + \bar{u}_b \sigma_{\mu\nu} C \bar{d}_a^T) \\ & + u_a^T
C \sigma_{\mu\nu} d_b (\bar{u}_a \gamma^\nu C \bar{d}_b^T +
\bar{u}_b \gamma^\nu C \bar{d}_a^T) \, ,
\end{array}
\\ &&
\begin{array}{ll}
\eta^A_{3\mu} & \sim u_a^T C \gamma_5 s_b (\bar{u}_a \gamma_\mu
\gamma_5 C \bar{s}_b^T -\bar{u}_b \gamma_\mu \gamma_5 C \bar{s}_a^T)
\\ & + u_a^T C \gamma_\mu \gamma_5 s_b (\bar{u}_a \gamma_5 C
\bar{s}_b^T - \bar{u}_b \gamma_5 C \bar{s}_a^T) \, ,
\\ \eta^A_{4\mu}
& \sim u_a^T C \gamma^\nu s_b (\bar{u}_a \sigma_{\mu\nu} C
\bar{s}_b^T + \bar{u}_b \sigma_{\mu\nu} C \bar{s}_a^T) \\ & + u_a^T
C \sigma_{\mu\nu} s_b (\bar{u}_a \gamma^\nu C \bar{s}_b^T +
\bar{u}_b \gamma^\nu C \bar{s}_a^T) \, ,
\end{array}
\end{eqnarray}
where $\eta^A_{1\mu}$ and $\eta^A_{2\mu}$ are the two independent
currents containing only light flavors; $\eta^A_{3\mu}$ and
$\eta^A_{4\mu}$ are the two independent ones containing one $s \bar
s$ pair.

\item For the isospin
singlet of $(\mathbf{\bar 3}_f \otimes \mathbf{\bar
6}_f)\oplus(\mathbf{6}_f \otimes \mathbf{3}_f)$ ($\mathbf{M}$),
\begin{eqnarray}\label{def:Mcurrent}
&&
\begin{array}{ll}
\eta^M_{1\mu} & \sim u_{a}^T C \gamma_\mu s_{b} (\bar{u}_{a} C
\bar{s}_{b}^T + \bar{u}_{b} C \bar{s}_{a}^T) \\ & + u_{a}^T C s_{b}
(\bar{u}_{a} \gamma_\mu C \bar{s}_{b}^T + \bar{u}_{b} \gamma_\mu C
\bar{s}_{a}^T) \, ,
\\ \eta^M_{2\mu} & \sim u_{a}^T C \sigma_{\mu\nu} \gamma_5
s_{b} (\bar{u}_{a} \gamma^{\nu} \gamma_5 C \bar{s}_{b}^T +
\bar{u}_{b} \gamma^{\nu} \gamma_5 C \bar{s}_{a}^T) \\ & + u_{a}^T C
\gamma^{\nu} \gamma_5 s_{b} (\bar{u}_{a} \sigma_{\mu\nu} \gamma_5 C
\bar{s}_{b}^T + \bar{u}_{b} \sigma_{\mu\nu} \gamma_5 C
\bar{s}_{a}^T) \, ,
\\ \eta^M_{3\mu} & \sim u_{a}^T C s_{b} (\bar{u}_{a} \gamma_\mu C
\bar{s}_{b}^T - \bar{u}_{b} \gamma_\mu C \bar{s}_{a}^T) \\ & +
u_{a}^T C \gamma_\mu s_{b} (\bar{u}_{a} C \bar{s}_{b}^T -
\bar{u}_{b} C \bar{s}_{a}^T) \, ,
\\ \eta^M_{4\mu} & \sim u_{a}^T C \gamma^{\nu} \gamma_5
s_{b} (\bar{u}_{a} \sigma_{\mu\nu} \gamma_5 C \bar{s}_{b}^T -
\bar{u}_{b} \sigma_{\mu\nu} \gamma_5 C \bar{s}_{a}^T) \\ & + u_{a}^T
C \sigma_{\mu\nu} \gamma_5 s_{b} (\bar{u}_{a} \gamma^{\nu} \gamma_5
C \bar{s}_{b}^T - \bar{u}_{b} \gamma^{\nu} \gamma_5 C \bar{s}_{a}^T)
\, ,
\end{array}
\end{eqnarray}
%%%%%%%%%%%%%%%%%%%%%%%%%%%%%%%%%%%%%%%%%%%%%%%%%%%%%%%%%%%%%%%%%%%%%%%%%%%%%%
%
where $\eta^M_{i\mu}$ are the four independent ones containing one
$s \bar s$ pair. The above structure has some implications on
their decay patterns.

\end{enumerate}
The expressions of Eqs.~(\ref{def:Scurrent})-(\ref{def:Mcurrent})
are not exactly correct, since they do not have a definite isospin.
For instance, the current $\eta^A_{3\mu}$ should contain $(u s \bar
u \bar s + d s \bar d \bar s)$ in order to have $I=0$. However, in
the following QCD sum rule analysis, we find that there is no
difference between these two cases in the limit that the masses and
condensates of the $up$ and $down$ quarks are the same. Actually we
also ignore a small quark mass effect ($m_u \sim m_d \lesssim 10$
MeV).

By using these tetraquark currents, we have performed the OPE
calculation up to dimension 12. Values for various condensates and
$m_s$ follow the references~\cite{Yao:2006px,values}. There are
altogether 14 currents. It turns out that some of them lead to the
same results of OPEs as the previous ones in
Ref.~\cite{Chen:2008qw}: $\eta^S_{1,2,3,4\mu} \sim
\eta^S_{1,2,3,4\mu}$~\cite{Chen:2008qw}, $\eta^A_{3,4\mu} \sim
\eta^A_{1,2\mu}$~\cite{Chen:2008qw}, and $\eta^M_{1,2,3,4\mu} \sim
\eta^M_{5,6,7,8\mu}$~\cite{Chen:2008qw}. Therefore, we just need
calculate the OPEs of $\eta^S_{5,6\mu}$ and $\eta^A_{1,2\mu}$. The
full OPE expressions are too lengthy and are omitted here.

In our previous paper~\cite{Chen:2008qw} we have found that the
OPEs of the currents $\eta^S_{i\mu}$'s and $\eta^A_{i\mu}$'s lead
to unphysical results where the spectral densities $\rho(s)$
become negative in the region of $2$ GeV$^2$ $\lesssim s \lesssim
4$ GeV$^2$. We find this to be the case also for the isoscalar
currents. Therefore, our QCD sum rule analysis does not support a
tetraquark state which has a flavor structure either $\mathbf{6_f}
\otimes \mathbf{ \bar 6_f }$ or $\mathbf{ \bar 3_f} \otimes
\mathbf{ 3_f }$ and a mass less than 2 GeV.

We shall discuss only the currents of the mixed flavor symmetry. We
find there is only one set of four independent currents as given in
Eqs.~(\ref{def:Mcurrent}), unlike the isovector case which have two
sets. The spectral densities calculated by the mixed currents are
positive for a wide range of $s$, and the convergence of OPE is very
good in the region of $2$~GeV$^2<M_B^2<$~5GeV$^2$ as in our previous
study~\cite{Chen:2008qw}. In general, the pole contribution should
be large enough in the SVZ sum rule. However, the pole contributions
of multiquark states are rather small due to the large continuum
contribution. Therefore a careful choice of the threshold parameter
is important in order to subtract the continuum contribution.
%At
%this moment we do not have a complete solution to this problem,
%while we can perform a sum rule analysis phenomenologically. Besides
%the SVZ sum rule, we will also use the finite energy sum rule. As we
%shall discuss in the following, the remarkable stability in both the
%SVZ sum rule and the finite energy sum rule indicates the signal of
%the physical state of the present exotic channel with a very similar
%mass.

\begin{figure}[!t]
\begin{center}
\scalebox{0.4}{\includegraphics{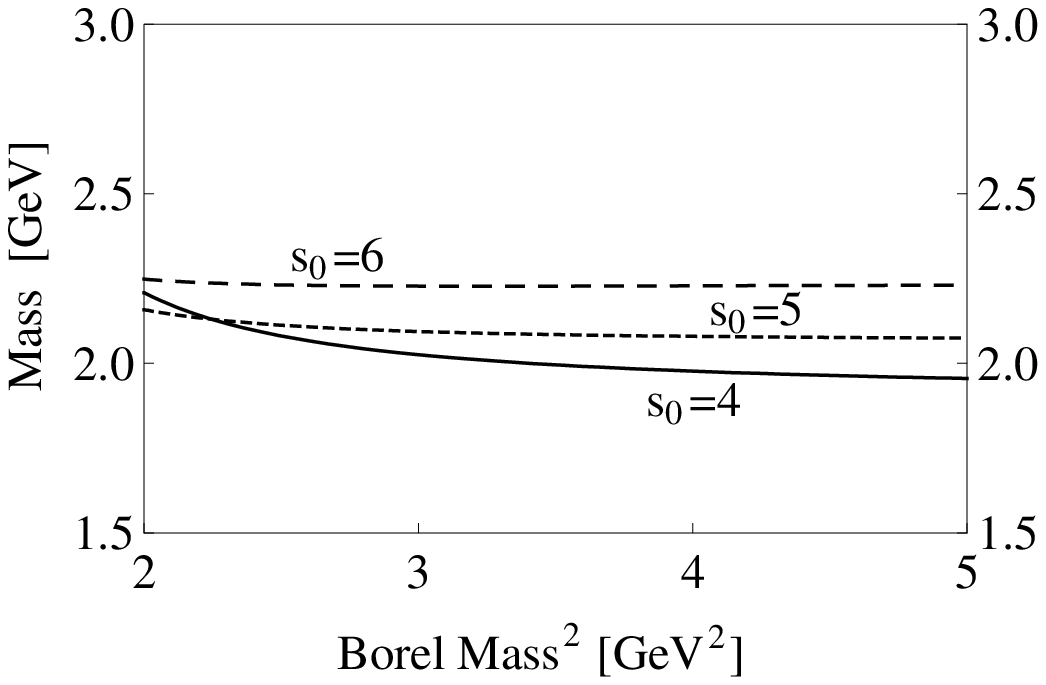}}
\scalebox{0.4}{\includegraphics{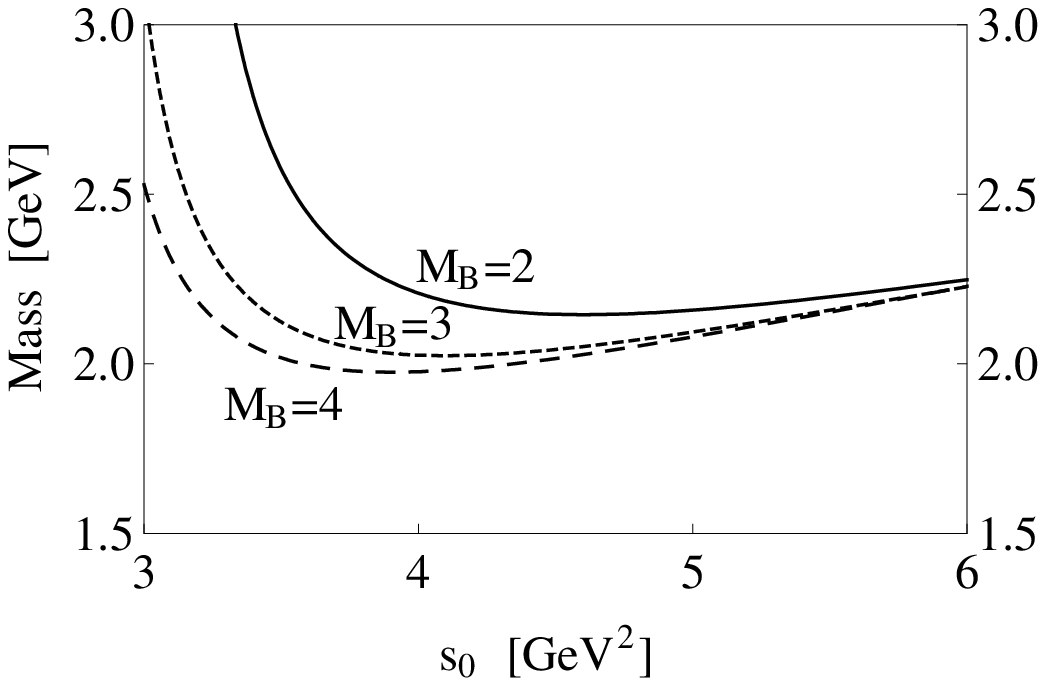}} \caption{The
mass of the state $q s \bar q \bar s$ calculated by using the
current $\eta^M_{2\mu}$, as functions of $M_B^2$ (left) and $s_0$
(right) in units of GeV.}\label{fig:eta2}
\end{center}
\end{figure}

When using the SVZ sum rule, the mass is obtained as functions of
Borel mass $M_B$ and threshold value $s_0$. As an example, we show
the mass calculated from currents $\eta^M_{2\mu}$ in
Fig.~\ref{fig:eta2}. The Borel mass dependence is weak, as shown
in the upper figure; the $s_0$ dependence has a minimum where the
stability is the best, as shown in the bottom figure. The minimum
is around 2.0 GeV, which we choose to be our prediction. The other
three independent currents $\eta^M_{1\mu}$, $\eta^M_{3\mu}$ and
$\eta^M_{4\mu}$ lead to similar results, which are around 2.1 GeV,
1.9 GeV and 2.0 GeV respectively.
%
%%%%%%%%%%%%%%%%%%%%%%%%%%%%%%%%%%%%%%%%%%%%%%%%%%%%%%%%%%%%%%%%%%%%%%%%%%%%%%
%%%%%%%%%%%%%%%%%%%%%%%%%%%%%%%%%%%%%%%%%%%%%%%%%%%%%%%%%%%%%%%%%%%%%%%%%%%%%%
%

\begin{figure}[h!t]
\begin{center}
\scalebox{0.5}{\includegraphics{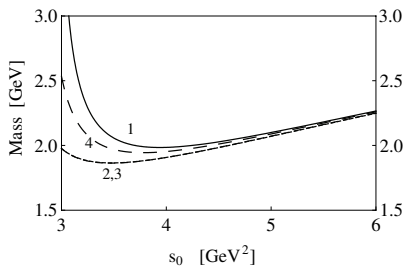}} \caption{The mass
calculated using the finite energy sum rule. The labels besides the
lines indicate the suffix i of the current $\eta^M_{i\mu}$
($i=1,\cdots,4$).} \label{fig:fesr}
\end{center}
\end{figure}

When using the finite energy sum rule, the mass is obtained as a
function of the threshold value $s_0$, which is shown in
Fig.~\ref{fig:fesr}. There is also a mass minimum around 2.1 GeV,
1.9 GeV, 1.9 GeV and 2.0 GeV for currents $\eta^M_{1\mu}$,
$\eta^M_{2\mu}$, $\eta^M_{3\mu}$ and $\eta^M_{4\mu}$ respectively.
In a short summary, we have performed a QCD sum rule analysis for $q
s \bar q \bar s$. The mass obtained is around 2.0 GeV. We label this
state $\sigma_1(2000)$.

%
%%%%%%%%%%%%%%%%%%%%%%%%%%%%%%%%%%%%%%%%%%%%%%%%%%%%%%%%%%%%%%%%%%%%%%%%%%%%%%
%---------figure 6*6
%%%%%%%%%%%%%%%%%%%%%%%%%%%%%%%%%%%%%%%%%%%%%%%%%%%%%%%%%%%%%%%%%%%%%%%%%%%%%%
%

\begin{figure}[h!t]
\begin{center}
\scalebox{0.5}{\includegraphics{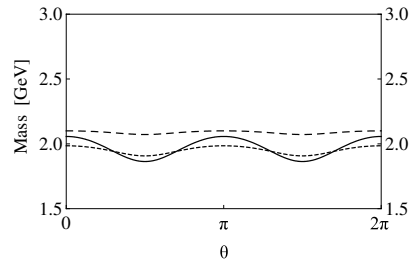}} \caption{The mass
calculated using the finite energy sum rule, and for the mixed
current $\eta^M_{mix}$. The curves are obtained by setting $s_0=3.5$
GeV$^2$ (solid line), 4 GeV$^2$ (short-dashed line) and 5 GeV$^2$
(long-dashed line).} \label{fig:mixing}
\end{center}
\end{figure}

We can also study the mixing of these four currents. The currents
$\eta^M_{1\mu}$ and $\eta^M_{3\mu}$ have the largest mass
difference, so we study their mixing as an example:
\begin{eqnarray}
\eta^M_{mix} = {\rm cos}\theta \eta^M_{1\mu} + {\rm sin}\theta
\eta^M_{3\mu} \, ,
\end{eqnarray}
where $\theta$ is the mixed angle. We calculate its OPE, and find
that the resulting spectral density is just:
\begin{eqnarray}
\rho^M_{mix} = {\rm cos}^2\theta \rho^M_{1\mu} + {\rm sin}^2\theta
\rho^M_{3\mu} \, ,
\end{eqnarray}
The obtained mass is shown in Fig.~\ref{fig:mixing} as functions of
$\theta$. When we take $s_0=3.5$ GeV$^2$ (solid line), the mass
maximum is 2.05 GeV, and the minimum is 1.85 GeV. Therefore, we
arrive at the similar result which produces the mass around 2 GeV.
We can also consider the mixing of other currents, which would not
change the results significantly due to the similarity of single
currents. The mass estimates are around $1.8 - 2.1$ GeV, where the
uncertainty is due to the mixing of two single currents.
%
%%%%%%%%%%%%%%%%%%%%%%%%%%%%%%%%%%%%%%%%%%%%%%%%%%%%%%%%%%%%%%%%%%%%%%%%%%%%%%
%---------figure 6*6
%%%%%%%%%%%%%%%%%%%%%%%%%%%%%%%%%%%%%%%%%%%%%%%%%%%%%%%%%%%%%%%%%%%%%%%%%%%%%%
%

Now let us discuss its decay properties as expected from a naive
fall-apart process. As shown in Eqs.~(\ref{def:Mcurrent}) the
currents contain one $s \bar s$ pair. Therefore, we expect that the
final states should also contain one $s \bar s$ pair. In order to
spell out the possible spin of decaying particles and their orbital
angular momentum, we need perform a Fierz rearrangement to change
$(qq)(\bar q \bar q)$ currents to $(\bar q q)(\bar q q)$ ones. For
illustration, we use one of the four independent $(\bar q q)(\bar q
q)$ currents~\cite{Chen:2008qw}:
%
%%%%%%%%%%%%%%%%%%%%%%%%%%%%%%%%%%%%%%%%%%%%%%%%%%%%%%%%%%%%%%%%%%%%%%%%%%%%%%
\begin{eqnarray}\label{def:mesonM}
%-------------------------------------xi M 2------------------------------------
\xi^{M}_{2\mu} &=& \nonumber (\bar{s}_{a} \gamma^\mu \gamma_5
s_{a})(\bar{u}_{b} \gamma_5 u_{b}) - (\bar{s}_{a} \gamma_5
s_{a})(\bar{u}_{b} \gamma^\mu \gamma_5 u_{b}) \\ && + \cdots \, .
\end{eqnarray}
%%%%%%%%%%%%%%%%%%%%%%%%%%%%%%%%%%%%%%%%%%%%%%%%%%%%%%%%%%%%%%%%%%%%%%%%%%%%%%
%
All terms of this current have the structure $(\bar{q}_{a}
\gamma^\mu \gamma_5 q_{a})(\bar{q}_{b} \gamma_5 q_{b})$.
Therefore, the expected decay patterns are: (1) $1^+$ and $0^-$
particles with relative angular momentum $L=0$, and (2) $0^-$ and
$0^-$ particles with $L=1$.

For the $S$-wave decay, we expect the following two-body decay
patterns
\begin{eqnarray}
\nonumber \sigma_1 (I^GJ^{PC} = 0^+ 1^{-+}) &\rightarrow& a_1(1260)
\eta, a_1 \eta^\prime, \cdots \, , \\ && b_1(1235) \eta, b_1
\eta^\prime \cdots \, .
\end{eqnarray}
If we consider, however, the $G$ parity conservation, the fist
line is forbidden and the second line is the only one allowed.
These modes can be observed in the final states $\omega \pi \eta$
and $\omega \pi \eta^\prime$.

For the $P$-wave decay, we expect (with the $G$ parity
conservation):
\begin{eqnarray}
\sigma_1 (I^GJ^{PC} = 0^+ 1^{-+}) &\rightarrow& KK, \eta \eta, \eta
\eta^\prime, \eta^\prime \eta^\prime \cdots \, .
\end{eqnarray}

We can also estimate the (partial) decay width through the
comparison with the observed $\pi_1(2015)$~\cite{Lu:2004yn}, which
has $\Gamma_{\rm tot} \sim 230$ MeV. Assuming that the decay of
$\pi_1(2015)$ solely goes through $S$-wave $b_1 \pi$ and that of
$\sigma_1(2000)$ through $b_1 \eta$, we expect
$\Gamma_{\sigma_1\rightarrow b_1 \eta} \sim 160$ MeV, as they are
proportional to the $S$-wave phase space. For the $P$-wave decay
there is an information $\pi_1(2015)\rightarrow\eta^\prime \pi$,
which corresponds to $\sigma_1(2000)\rightarrow\eta^\prime \eta$
(Because both $\pi_1(1600)$ and $\pi_1(2015)$ have been observed in
the final states $\pi \eta^\prime$ other than $\pi \eta$, we choose
$\eta \eta^\prime$ to be the final states of $\sigma_1(2000)$ other
than $KK$ and $\eta \eta$). Assuming once again that this is the
unique decay mode, we expect that the decay width is approximately
130 MeV. If the decay occurs $50\%$ through $b_1 \pi$ ($b_1 \eta$)
and $50\%$ through $\eta^\prime \pi$ ($\eta^\prime \eta$), we expect
that $\Gamma_{\sigma_1} \sim 150$ MeV.

In summary, we have performed the QCD sum rule analysis of the
exotic tetraquark states with $I^GJ^{PC} = 0^+1^{-+}$. We test all
possible flavor structures in the diquark-antidiquark $(qq)(\bar q
\bar q)$ construction, $\mathbf{6} \otimes \mathbf{\bar 6}$,
$\mathbf{\bar 3} \otimes \mathbf{3}$ and $(\mathbf{\bar 3} \otimes
\mathbf{\bar 6}) \oplus (\mathbf{6} \otimes \mathbf{3})$. We find
that only the mixed currents of the flavor structure $(\mathbf{\bar
3} \otimes \mathbf{\bar 6}) \oplus (\mathbf{6} \otimes \mathbf{3})$
allow a positive and convergent OPE, and there is only one choice
with the quark content $q s \bar q \bar s$, which have four
independent currents. We have then performed both the SVZ sum rule
and the finite energy sum rule. The mass estimates are around $1.8 -
2.1$ GeV, where the uncertainty is due to the mixing of two single
currents. The possible decay modes are $S$-wave $b_1(1235) \eta$ and
$b_1(1235) \eta^\prime$, and $P$-wave $KK$, $\eta \eta$, $\eta
\eta^\prime$ and $\eta^\prime \eta^\prime$, etc. The decay width is
around 150 MeV through a rough estimation. Here we want to note that
we do not know how to determine the mixing angle, which is an
interesting problem.

%
%=====================================================================================
%=====================================================================================
%=====================================================================================
\section*{Acknowledgments}
%=====================================================================================
%=====================================================================================
%=====================================================================================
%

H.X.C. is grateful for Monkasho support for his stay at the Research
Center for Nuclear Physics where this work was done. This project
was supported by the National Natural Science Foundation of China
under Grants No.~10625521, 10721063, the Ministry of Education of
China, and the Grant for Scientific Research ((C) No.~19540297) from
the Ministry of Education, Culture, Science and Technology, Japan.

%%%%%%%%%%%%%%%%%%%%%%%%%%%%%%%%%%%%%%%%%%%%%%%%%%%%%%%%%%%%%%%%%%%%%%%%%%%%%%

%%%%%%%%%%%%%%%%%%%%%%%%%%%%%%%%%%%%%%%%%%%%%%%%%%%%%%%%%%%%%%%%%%%%%%%%%%%%%%
%

\end{document}